\documentclass[prd,aps,preprint,groupedaddress,superscriptaddress,nofootinbib]{revtex4}

\usepackage{comment}
\usepackage[utf8]{inputenc} 
\usepackage{graphicx,color,overpic,mathtools}
\usepackage{amsthm,amsmath,amssymb,hyperref,mathrsfs}
\usepackage{braket,bm,bbm,setspace}
\PassOptionsToPackage{normalem}{ulem}
\usepackage{ulem} 
\usepackage{physics}
\usepackage{float}
\usepackage[makeroom]{cancel}
\usepackage[english]{babel}
\usepackage{graphicx}
\usepackage{xcolor}
\graphicspath{ {images/} }
\addto\captionsspanish{}
\hypersetup{
    colorlinks=false,
    pdfborder={0 0 0},
}

\definecolor{rRGB}{RGB}{255, 0, 255}

\begin{document}

\hfill YITP-22-154
\vspace*{1.4cm}
\title{Comment on ``Regular evaporating black holes with stable cores"}
\vspace*{-1.5cm}
\author{Ra\'ul Carballo-Rubio}
\affiliation{CP3-Origins, University of Southern Denmark, Campusvej 55, DK-5230 Odense M, Denmark}
\affiliation{Florida Space Institute, University of Central Florida, 12354 Research Parkway, Partnership 1, 32826 Orlando, FL, USA}
\author{Francesco Di Filippo}
\affiliation{Center for Gravitational Physics, Yukawa Institute for Theoretical Physics, Kyoto University, Kyoto 606-8502, Japan}
\author{Stefano~Liberati}
\affiliation{SISSA - International School for Advanced Studies, Via Bonomea 265, 34136 Trieste, Italy}
\affiliation{IFPU, Trieste - Institute for Fundamental Physics of the Universe, Via Beirut 2, 34014 Trieste, Italy}
\affiliation{INFN Sezione di Trieste, Via Valerio 2, 34127 Trieste, Italy}
\author{Costantino Pacilio}
\affiliation{Dipartimento di Fisica “G. Occhialini”, Università degli Studi di Milano-Bicocca, Piazza della Scienza 3, 20126 Milano, Italy}
\affiliation{INFN, Sezione di Milano-Bicocca, Piazza della Scienza 3, 20126 Milano, Italy}
\author{Matt Visser}
\affiliation{
School of Mathematics and Statistics, Victoria University of Wellington, PO Box 600, Wellington 6140, New Zealand
}

\begin{abstract}
Regular black holes are generically unstable because of the classical phenomenon that goes by the name of ``mass inflation" which destabilizes the inner horizon. In a recent article, 
[Phys.Rev.D~107~(2023)~2,~024005], it is argued that semiclassical effects due to Hawking radiation can cure this instability, and some concerns are raised against the validity of previous analyses showing its existence in the first place. In this short comment, we explain our reservations regarding these recent claims, and reiterate the relevance of the mass inflation instability for regular black holes of astrophysical interest.
\end{abstract}


\maketitle

In a recent work~\cite{Bonanno:2022Hawking}, it is argued that the backreaction of Hawking evaporation is a relevant factor in the analysis of the stability properties of regular black holes. However, in this comment we show that this claim is based on an extrapolation of the mathematical model used beyond its regime of validity.

The model used in~\cite{Bonanno:2022Hawking} is an extension of the well-known Ori model~\cite{Ori:1991PRL,Levin:1996PRD}, used to study the instability of Reissner-Nordstr\"om black holes. This extension can be applied to study the behavior of different models of regular black hole. Once a specific regular black hole model is chosen, the following results follow~\cite{Carballo-Rubio:2021JHEP} (see also~\cite{DiFilippo:2022Universe} for a complementary discussion):
\begin{itemize}
    \item Regardless of the model selected, there is an exponential mass inflation of the Misner--Sharp mass that continues until backreaction cannot be ignored, and thus the linear approximation breaks down (in complete parallelism with the Ori model~\cite{Ori:1991PRL}).
    \item If one insists in using the (unphysical) linear approximation beyond this point, then the Misner--Sharp mass is always divergent but different background geometries can display different degrees of divergence (ranging from exponential to polynomial). 
\end{itemize}
While from a mathematical perspective it is always possible to push the linear model to its ultimate consequences, such an exercise has clearly no physical relevance.

The authors of~\cite{Bonanno:2022Hawking} call this polynomial behavior, in the models in which it appears, ``late-time dynamics''. For instance, in page 2 of~\cite{Bonanno:2022Hawking}, the authors write:

\begin{quote}
\textit{``In [30], it was established that this conclusion is
premature though. While the extrapolation works for certain classes of static regular black holes, including the geometries proposed by Bardeen, the Hayward geometry and renormalization group improved black hole solutions are free from mass inflation. In these cases the mass function at the Cauchy horizon grows polynomially in time only and the resulting curvature singularity may be integrable. Technically, this behavior can be traced back to the presence of a late-time attractor in the evolution equation for the mass-function at the Cauchy horizon, rendering this quantity finite at asymptotically late times."}
\end{quote}
The statement above that some classes of regular black holes are free from mass inflation is not correct. As mentioned above and elaborated below, that the late-time behavior is not exponential does not mean that mass inflation has not taken place, and is in any case outside of the rabge of validity of the Ori model.

To make this concrete, let us consider the behavior of the Misner-Sharp mass explicitly. If the Misner--Sharp mass depends on the asymptotic mass linearly, \textit{i.e.}, 
\begin{equation}\label{eq:M-linear}
    M(v,r)=g_1(r)\,m(v)+g_2(r),
\end{equation}
then the behavior of the Misner--Sharp mass is always exponential,
\begin{equation}
    M_+\sim \frac{e^{|\kappa_-|v}}{v^{\gamma+1}}\,,
\end{equation}
where $\kappa_-$ is the surface gravity of the inner horizon. Geometries in this class include the Reissner--Nordstr\"om black hole and the Bardeen regular black hole \cite{Bardeen:1968}. 

On the other hand, for Hayward's regular black hole \cite{Hayward:2005}, after the initial exponential phase, and beyond the regime of validity of the linear approximation, we get power-law behaviour
\begin{equation}\label{eq:Poly}
  M_+\propto \left|\kappa_-\right|\frac{v^{\gamma+1}}{\beta}\,. 
\end{equation}
One could think that this different polynomial behavior could alleviate the instability, in particular when Hawking radiation is taken into account; this is what motivates the authors of~\cite{Bonanno:2022Hawking}. However, by the end of the exponential phase the backreaction on the geometry is already very large, and the linear approximation at the core of the model can no longer be trusted. In the case of Hayward's metric, the transition between the exponential and polynomial phase occurs when the ratio, between the Misner--Sharp mass in the interior region $M_+$, and the initial mass $m_0$, is given by
\begin{equation}\label{eq:transition}
    \frac{M_+}{m_0}\sim \,\frac{v^{\gamma+1}}{6\beta}\,\frac{m_0}{\ell}\gg1\,.
\end{equation}
Hence, the backreaction of perturbations cannot be ignored, and the linear model ceases to be valid. Any conclusions that result from the application of this linear model beyond its regime of validity are not reliable.

In particular, this applies to the analysis of the impact that Hawking evaporation has on this ``late-time behavior'' that is the driving motivation in~\cite{Bonanno:2022Hawking}:

\begin{quote}
\textit{``The taming of the mass-inflation effect then suggests that the dynamics at the Cauchy horizon and the black hole evaporation process could happen on similar timescales. Thus, a more complete understanding of the actual dynamics mandates to take the Hawking evaporation process into account. Our work addresses this question for the first time. As our main result, we discover two classes of universal late-time behaviors whose properties are dictated by simple structural properties of the mass function and the universality of the Hawking effect. The late-time attractors governing the dynamics either lead to a polynomial growth of the squared curvature tensors or even renders these quantities finite. Interestingly, the latter behavior appears for the case of Reissner-Nordstr\"{o}m geometry once the Hawking radiation is included. The final state of the black hole evaporation process is still a cold remnant.''}
\end{quote}



Note also that the authors of~\cite{Bonanno:2022Hawking} are not properly appreciating the extremely important separation of scales between ringdown, the Price regime, and the truly enormous timescale associated with backreaction induced by Hawking radiation. 

\begin{quote}
\textit{``We stress that, strictly speaking, the Ori model building on the Price tail behavior (10) is valid for asymptotically late times $v$ only. Therefore, conclusions drawn from the model in a regime where $v$ is small and the perturbation significant compared to the mass $m_0$ have to be interpreted with care and should be confirmed by an analysis of the full dynamics [36].''}
\end{quote}
What the current authors call ``late times"  means late enough such that the Price law provides a good description of the system. It is true that at very early times after the black hole spacetime is perturbed Price's law cannot be applied, but after a time scale $v\sim \mathcal{O}(M)$ (corresponding to the ringdown time) it becomes the dominant one, while Hawking radiation is completely negligible for a much longer timescale $v\sim \mathcal{O}(M^3/M_{Planck}^2)$. Due to this vast separation of scales, there is a very long transient in which the exponential phase in the Ori model discussed in~\cite{Carballo-Rubio:2022JHEP} is building up. 

Let us also stress that while an idealized analysis of the role of Hawking radiation may be legitimate, still one should not neglect the CMB role when the analysis is aimed at testing the phenomenological relevance of regular black holes. Indeed, for a solar mass black hole, Hawking radiation is subdominant even with respect to the incoming CMB radiation. Therefore, after the decay of the perturbations following Price's law, there will be an extremely long transient in which the dominant perturbation is given by incoming CMB radiation, not by outgoing Hawking radiation. For a solar mass black hole, the order of magnitude of such a transient can be easily estimated to be around 300 billion years~\cite{Taylor:2022pgs}, and is hence even longer for supermassive black holes, which gives plenty of time for the exponential buildup and subsequent destabilization of the core. 
Nonetheless, still this should not overshadow the main message of this comment that the early exponential phase we discussed above is present even if the CMB is ignored.


For completeness, let us stress that our analysis should not be taken as a criticism to models predicting regular black holes as a possible piece in the resolution of the singularity problem. In fact, our analysis simply shows that if such solutions are generated, a small perturbation has a large backreaction on the geometry  which, as a consequence, will then become non-stationary. The investigation concerning the end point of such evolution is a question that should be addressed on a  case by case basis and which depends on the dynamical equations of the theory. 

Possible scenarios at the moment entail an outcome in a regular black hole configuration with zero surface gravity at the inner horizon~\cite{Carballo-Rubio:2022JHEP, Franzin:2022wai}, the increase of the regularization length $\ell$ so to generate a horizonless configuration~\cite{Barcelo:2022gii,Carballo-Rubio:2022RBH2HCO}, a series of bounces that result into the formation of a horizonless configuration~\cite{Barcelo:2014cla,Barcelo:2016hgb}, or a bounce destroying the trapped region~\cite{Haggard:2014rza,Husain:2021ojz} (albeit in the latter case, the mechanism that would allow to reconcile the instability timescale with the long enough timescales needed to accommodate observations of astronomical black holes has not been identified). All of these possibilities lead to interesting phenomenological consequences which are worth exploring~\cite{Carballo-Rubio:2018jzw,Cardoso:2019rvt,Eichhorn:2022oma,Carballo-Rubio:2022imz,Eichhorn:2022fcl,Carballo-Rubio:2022aed}. Less so, effectively eternal regular black holes.

\section*{Acknowledgements}
RCR acknowledges financial support through a research grant (29405) from VILLUM fonden.
FDF acknowledges financial support by Japan Society for the Promotion of Science Grants-in-Aid for international research fellow No. 21P21318. 
SL acknowledges funding from the Italian Ministry of Education and  Scientific Research (MIUR)  under the grant  PRIN MIUR 2017-MB8AEZ. 
CP is supported by European Union's H2020 ERC Starting Grant No.~945155-GWmining and Cariplo Foundation Grant No.~2021-0555.
MV was supported by the Marsden Fund, via a grant administered by the Royal Society of New Zealand. 

\clearpage
\bibliography{refs}
\end{document}